# Tellurium-bridged two-leg spin ladder in Ba$_2$CuTeO$_6$

G. Narsinga Rao,[1] R. Sankar,[1,2] Akansha Singh,[3] I. Panneer Muthuselvam,[1] W.T. Chen,[1] Viveka Nand Singh,[4,5] Guang-Yu Guo,[4,6,*] and F. C. Chou[1,7,8,*]

[1]Center for Condensed Matter Sciences, National Taiwan University, Taipei 10617, Taiwan
[2]Institute of Physics, Academia Sinica, Taipei 10617, Taiwan
[3]Harish-Chandra Research Institute, Allahabad 211019, India
[4]Department of Physics, National Taiwan University, Taipei 10617, Taiwan
[5]Institute of Atomic and Molecular Sciences, Academia Sinica, Taipei 10617, Taiwan
[6]Physics Division, National Center for Theoretical Sciences, Hsinchu 300, Taiwan
[7]National Synchrotron Radiation Research Center, Hsinchu 30076, Taiwan
[8]Taiwan Consortium of Emergent Crystalline Materials, Ministry of Science and Technology, Taipei 10622, Taiwan

We present single-crystal growth and magnetic property studies of tellurium-bridged copper spin-1/2 system Ba$_2$CuTeO$_6$. The spin-exchange interaction among copper spins via Cu-O-Te-O-Cu super-superexchange route leads to a novel two-leg spin ladder system. Spin susceptibility $\chi(T)$ data indicates that the triclinic Ba$_2$CuTeO$_6$ undergoes a stepwise crossover for exchange couplings revealed by a broad maximum near $T_{max}$~75 K and an anisotropic cusp in d$\chi$/d$T(T)$ at $T_N$~15 K to signify a three dimensional (3D) antiferromagnetic long-range ordering (LRO). The 3D LRO has been suggested from the anisotropic behavior of $\chi(T)$ with strong $c$-axis spin anisotropy and the signature of spin flop transition from the isothermal magnetization below $T_N$. Analysis of magnetic heat capacity ($C_m$) at $T_N$~15 K indicates that most of the spin entropy (~92 %) has already been released above $T_N$, which supports the picture of consecutive spin entropy reduction upon cooling with Te-bridged two-leg spin ladder system with strong intraladder and interladder couplings. Theoretical DFT+$U$ calculations have been performed to search for the ground state magnetic configuration and also to evaluate exchange coupling constants that support the magnetic model deduced from the combined spin susceptibility and crystal structure symmetry analysis.

## I. Introduction

Low-dimensional magnetism of the copper-based quantum spin-1/2 system is the foundation to understand high $T_c$-superconductivity. [1] While most of the high-$T_c$ cuprate superconductors contain electron or hole-doped $CuO_2$ planes of antiferromagnetic (AF) coupling in the normal state, such as the $YBa_2Cu_3O_7$ of $T_c$~93 K, [2] the resonant valence bond (RVB)-type bonding mechanism and the Cu-O-Cu spin superexchange (SE) coupling have been examined through many comparative studies of cuprate with various crystal and spin structures. [3-5] The spin structures have been analyzed from one dimensional (1D) chain to two dimensional (2D) plane extensively based on the Heisenberg and t-J models. [6] For example, the $Sr_2CuO_3$ with Cu-O network of corner-sharing chain, [7-9] the $CuGeO_3$ with edge-sharing chain, [10] the $Sr_{0.4}Ca_{13.6}Cu_{24}O_{41}$ with two-leg spin ladder,[11] the $SrCu_2O_3$ and $Sr_2Cu_3O_5$ with even- and odd-leg spin ladders,[12, 13] and the multiple number of $CuO_2$ planes per unit cell. [14] Following the development of W- or Te-bridged transition-metal (M) oxides containing spin chain and plane systems, such as $Li_2M(WO_4)_2$ and $A_2MBO_6$, where A = Sr or Ba, and B = Te or W, [15-18] the original SE interaction route of M-O-M has been modified via a Cu-O-(W,Te)-O-Cu of super-superexchange (SSE) interaction route. If two compounds of identical structure but of different SE and SSE origins, which may not be distinguishable within the theoretical Heisenberg Hamiltonian description, it would be interesting to have a chance to compare the impact of SSE on the spin-coupling forms of spin dimerization, AF ordering, and superconductivity.

Tellurates, oxides containing the $Te^{6+}$ cation, can often be utilized to build three-dimensional crystal structures attributable to their preferred octahedral coordination environment and their ability to bond a large number of metal centers, [19-21] as a result, tellurates rarely form low-dimensional crystal structures. In contrast, cuprate compounds have a rich variety of coordination environments, including square planar, square pyramidal, and tetragonally distorted octahedral coordinations. As a result, tellurium-bridged cuprate is a new class of material with a higher degree of freedom on both crystal and magnetic structure variations. Among the few reported examples of tellurium-bridged cuprate compounds, e.g., oxides containing $d^9$-$Cu^{2+}$ and $Te^{6+}$ cations, such as $Sr_2CuTeO_6$, [22] $Ba_2CuTeO_6$, [17] $Na_2Cu_2TeO_6$, [23, 24] $Tl_4CuTeO_6$, and $Tl_6CuTe_2O_{10}$, [25] all exhibit three-dimensional crystal structures by using Te as the bridging ion for the spin coupling in SSE form. Interestingly, $Na_2Cu_2TeO_6$ shows Cu spin dimers that are bridged by the $TeO_6$ octahedra, i.e., the Cu spin dimer has Cu-O-Cu of SE mechanism, but the inter-dimer coupling is of SSE mechanism. The mixed SE and SSE mechanisms for the tellurium-bridged cuprate compounds open up a new territory in the study of low dimensional magnetism.

$Ba_2CuTeO_6$ compound exhibits two phases, one is prepared at ambient pressure and the other can only be synthesized under high-pressure. Sample prepared at 900°C under 5 GPa pressure crystallizes in a perovskite-type structure of tetragonal distortion (Fig. 1(a)),[17] which shows a broad peak of $\chi(T)$ near 175 K to indicate a short-range antiferromagnet coupling without detectable three-dimensional (3D) long-range magnetic ordering down to ~2 K. The short-range AF coupling was attributed to the cooperative Jahn − Teller distortion of $CuO_6$ octahedra. On the other hand, the ambient pressure form crystallizes in a triclinic structure (Fig. 1(b)). [26] In the triclinic $Ba_2CuTeO_6$, the distortion of the $CuO_6$ octahedra is small compared to that of the high-pressure perovskite phase. A detailed study on the triclinic $Ba_2CuTeO_6$ is desirable to learn more about the dimensionality and the role of bridging Te in the $S =1/2$ quantum spin systems. In addition, physical properties of the triclinic $Ba_2CuTeO_6$ have not been reported so far, not to mention that using a single-crystal sample.

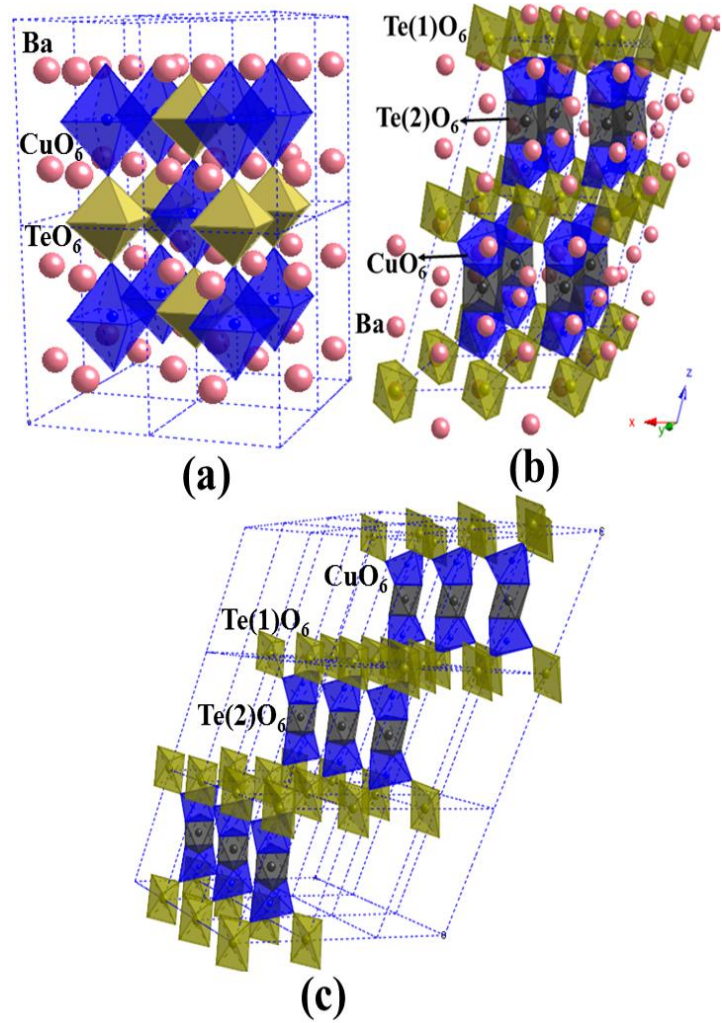

Figure 1: (Color online) (a) Crystal structure of tetragonal $Ba_2CuTeO_6$ high pressure phase. (b) Crystal structure of triclinic $Ba_2CuTeO_6$ is shown with $CuO_6$ (Blue), $Te(1)O_6$ (Yellow) and $Te(2)O_6$ octahedra (Black) stacked in layers and the Ba atoms are big sphere (Marron). (c) Single layer of a two-leg spin ladder is illustrated in triclinic $Ba_2CuTeO_6$, where $Te(2)O_6$ and $Te(1)O_6$ bridge two $CuO_6$ octahedra via face-sharing and corner-sharing of oxygen, respectively.

In this paper, we present the investigation of the thermodynamic and magnetic properties of triclinic $Ba_2CuTeO_6$ with single-crystal samples. Based on strong SSE coupling of $Cu^{2+}$ spins within the *ab* plane, short-range AF exchange correlations were found to appear as indicated by the existence of a broad peak of $\chi(T)$ at $T_{max} \sim$ 75 K. Moreover, a long-range AF-like anomaly of $T_N$ = 15 K has been identified from the cusp in the $d\chi/dT$ plot. These characteristic anomalies were also confirmed in $C_p$ measurements. Based on the unique geometric coordination between $CuO_6$ and $TeO_6$ octahedra, we find that these signatures of magnetic coupling shown in the experimental results of $\chi(T)$ and $C_p(T)$ could be attributed to the consecutive short-range AF spin-exchange couplings from the intrachain and interchain interactions of a two-leg spin ladder with nontrivial inter-ladder interactions, which eventually falls to a 3D long-range AF ordering of Cu spins below $T_N$. These experimental results were found to be consistent with the theoretical calculations within the framework of the density functional theory (DFT).

## II  Experimental and computational details

For the first step of the single crystal growth, the $Ba_2CuTeO_6$ polycrystalline sample was prepared by the solid-state reaction method. Stoichiometric proportions of high purity $BaCO_3$, CuO and $TeO_2$ powders were mixed and fired in the air at 1000°C for 12 h with a heating and cooling rate of 120 °C/h. The pre-heated powders were well grounded and re-heated at 1100 °C for 24 h with several intermediate grindings to reach single phase. The $Ba_2CuTeO_6$ single crystal was grown with flux method using $BaCl_2$ as the flux. The mixture of the polycrystalline sample of $Ba_2CuTeO_6$ and the flux of $BaCl_2$ in molar ratio of 1:3 was melted in an alumina crucible at 1150°C for 24 h. The furnace was slowly cooled to 850°C at the rate of 3°C /h and then cooled down to room temperature at the rate of 80°C/h. Dark green crystals (1-3 mm) were mechanically separated from the crucible and further washed with hot water. The crystal structure and phase purity of the samples were checked by synchrotron X-ray powder diffraction (SXRD) using an incident wavelength of $\lambda$ = 0.619927 Å (BL01C2, NSRRC, Taiwan). The field cooled (FC) and zero field cooled (ZFC) magnetization curves were measured in a commercial Vibrating Sample Magnetometer (VSM, Quantum Design, USA) from 1.8 K to 350 K in the presence of various applied magnetic fields. The isothermal magnetization ($M$) data were also recorded at selected temperatures. The heat-capacity ($C_p$) measurements were carried out by a relaxation method using the Physical Properties Measurement System (PPMS, Quantum Design, USA).

All theoretical calculations were performed within the framework of the density functional theory (DFT). Interaction between the valence electrons and the ion cores is represented by the projector augmented wave (PAW) potentials [27] as implemented in the Vienna *ab initio* simulation package (VASP). [28, 29] The generalized gradient approximation (GGA) [30] is used for the exchange-correlation functional. The wave functions were expressed in a plane wave basis set with an energy cutoff of 500 eV and the self consistent field energies are converged up to $10^{-6}$ eV. In order to describe the electron-electron correlation associated with the $3d$ states of Cu, the GGA plus on-site repulsion (GGA + $U$) [31] calculations are carried out with an effective $U_{eff}= (U-J) = 3.6$ eV. We note that LiCu$_2$O$_2$ as a low-dimensional AF oxide with frustrated exchange couplings, the $U_{eff} = (U-J) = 3.6$ eV used in the previous GGA +$U$ calculation is in good agreement with that extracted from the x-ray absorption experiment [32]. Therefore, we used the same $U_{eff}$ for Cu atoms in the present GGA+$U$ calculations for Ba$_2$CuTeO$_6$.

The crystal structure data for the *ab initio* calculations were taken from the refined lattice parameters. The primitive unit cell of Ba$_2$CuTeO$_6$ contains two formula units, *i.e.,* there are two Cu atoms per unit cell. In order to investigate the magnetic ground state of this system, we have considered a $(2\times2\times2)$ supercell. In the present calculations, we used the tetrahedron method with Blöchl corrections for the Brillouin zone integration with a $\Gamma$-centered Monkhorst-Pack k-point mesh of $(6\times6\times3)$. Further test calculations using denser *k*-point meshes and larger kinetic energy cutoffs showed that the calculated total energy differences of the magnetic states considered with respect to the ferromagnetic state converged well to within $10^{-4}$ eV.

To ensure that the structural parameters from the present theoretical calculations are not significantly different from the experimental ones, we have also determined both the lattice constants and atomic positions theoretically. First, we calculated the total energy for several sets of lattice constants and fit them to a volume $(V)$ polynomial of total energy $E = a_0 + a_1 V + a_2 V^2 + a_3 V^3$. The thus-obtained theoretical lattice constants are quite close to the experimental ones (being about 2% larger). Then, we optimized the atomic structure using the theoretical lattice constants. In the structural optimization, atoms are allowed to relax until the forces on the atoms are smaller than 0.01 eV/Å. We found that the calculated bond lengths and bond angles differ from the corresponding experimental values by only a few percent. Therefore, to have better agreement with the experiments, we present only the results of our *ab initio* calculations using the refined experimental structural parameters in this paper.

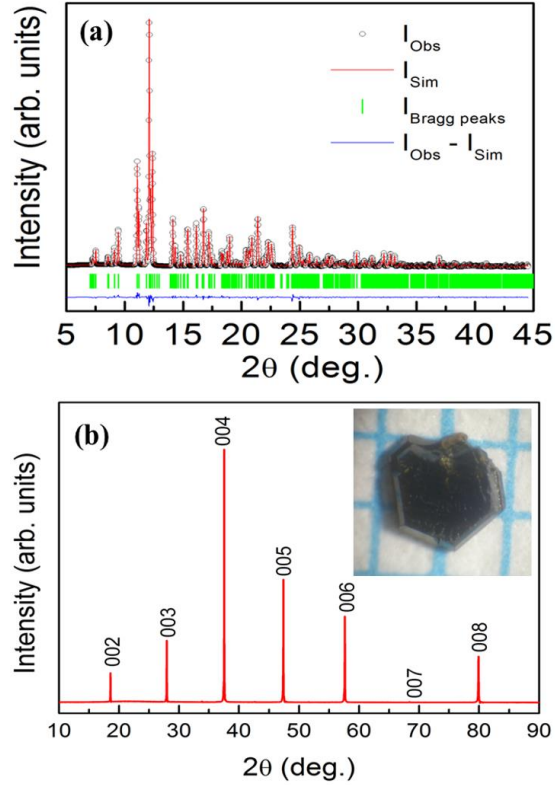

Figure 2: (Color online) (a) Room-temperature powder SXRD pattern of $Ba_2CuTeO_6$, where black circles are experimental data, solid curve in red is the best fit from the Rietveld refinement. The vertical bars indicate the position of Bragg peaks of space group $P\bar{1}$ and the bottom curve shows the difference between the observed and calculated intensities. (b) Single-crystal diffraction pattern obtained using an x-ray (Cu-K$_\alpha$) beam perpendicular to the *ab*-plane. Inset shows a photograph of grown single crystal.

### III. Results and discussion

#### A. Crystal Structure

The powder SXRD pattern of the polycrystalline $Ba_2CuTeO_6$ is shown in Fig. 2(a). The SXRD pattern can be indexed with a space group of $P\bar{1}$ in triclinic symmetry without any observable trace of impurity phases. The structural parameters were refined by the General Structure Analysis System (GSAS) program [33] following the Rietveld technique of satisfactory quality as indicated by the $R_{wp}$= 5.06 % and $R_p$= 3.79 %. The fitted lattice parameters are $a = 5.7288(1)$ Å, $b = 5.8677(1)$ Å, and $c = 10.2237(2)$ Å, $\alpha = 107.867(1)°$,

$\beta = 106.208(2)°$, and $\gamma = 60.750(2)°$, which are in good agreement with previously reported values. [26] The single crystal diffraction pattern shows only (00$l$) reflections (Fig. 2(b)), indicating the preferred $ab$-plane orientation of the as-grown crystal. Single crystals were obtained in the form of hexagonal geometric shapes as shown in the inset of Fig. 2(b). The ordered Ba$_2$CuTeO$_6$ could be viewed consisting of CuO$_6$ octahedra which are bridged through TeO$_6$ octahedra in either face-sharing (for Te(2)) or corner-sharing (for Te(1)), as shown in Fig. 1(b). Alternatively, Ba$_2$CuTeO$_6$ can also be viewed composing of Te(2)-bridged Cu dimers through the SSE route, as illustrated in Fig. 1(c).

## B. Magnetic susceptibility

Figure 3 shows the homogeneous magnetic susceptibility ($\chi = M/H$) of single-crystal Ba$_2$CuTeO$_6$ as a function of temperature $\chi(T)$ measured in an applied magnetic field of 10 kOe along ($H\|$) and perpendicular ($H\perp$) to the $ab$ plane, which shows an approximation to the actual $c$ axis when the actual crystal symmetry is triclinic of pseudo-hexagonal shape (inset of Fig. 2(b)) with $\alpha$ and $\beta$ angles are not in right angles but close to 100°. There was no detectable difference between data taken through the field-cooled (FC) and zero-field-cooled (ZFC) routes. As the temperature was lowered, an isotropic broad maximum of $\chi(T)$ at $T_{max}\sim75$ K was observed, presumably due to an AF-like short-range exchange correlations. Below about ~15 K, $\chi(T)$ measured at 10 kOe data were found anisotropic, which was confirmed with a low field of 100 Oe to indicate the onset of an AF-like long range ordering (LRO), as shown in the lower inset of Fig. 3(a).

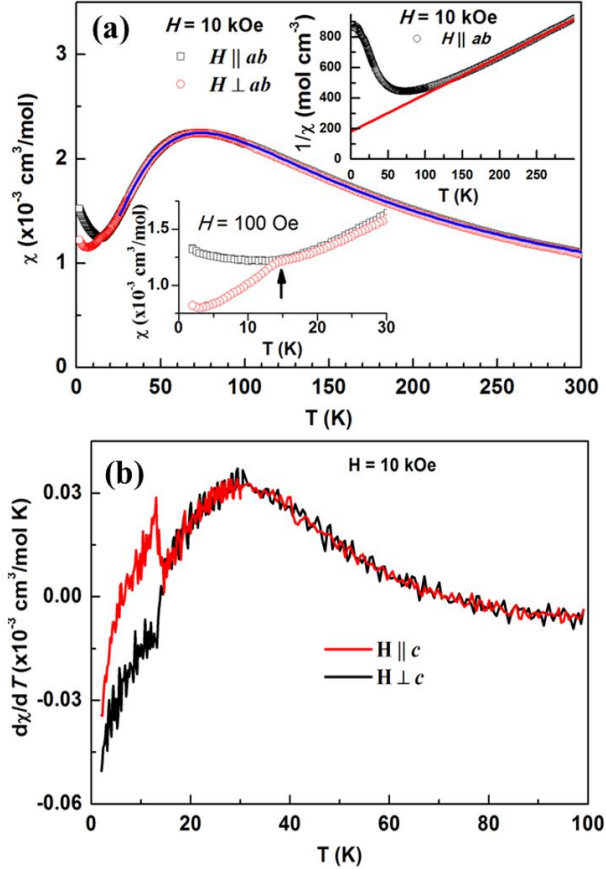

Figure 3: (Color online) (a) The temperature dependence of magnetic susceptibility measured in an applied magnetic field of 10 kOe for $H\|ab$ and $H\perp ab$ of $Ba_2CuTeO_6$ single crystal. The solid curve in blue is the best fit from the modified Bonner-Fisher AFM chain model. Upper inset shows $1/\chi(T)$ fitted to the Curie-Weiss law (solid red line). The low field $\chi(T)$ curve below 30 K is displayed in the lower inset with a transition temperature near ~15 K as indicated by the arrow. (b) The $d\chi/dT$ vs $T$ curves measured with field of 10 kOe reveal an anisotropic cusp of $T_N$ ~15 K.

By taking the temperature derivative of $\chi(T)$ measured at 10 kOe, as shown in Fig. 3(a), the $d\chi/dT$ $(T)$ curves reveal more distinct anomaly below $T_{max}$~75 K, a cusp of $d\chi/dT$ that suggests a sharp drop of $\chi(T)$ near $T_N$~15 K, as illustrated more clearly in Fig. 3(b). The broad maximum of $\chi(T)$ at $T_{max}$~75 K indicates the existence of a short range AF exchange correlations, as commonly found in the Cu-O chain or plane systems of superexchange spin coupling.[34, 35] While the d$\chi$/d$T$ cusp corresponds to the anisotropic sharp drop of $\chi(T)$, $T_N$ ~15 K strongly suggests the onset of a long range AF-like spin ordering with spins oriented mostly perpendicular to the *ab* direction, either as an AF or a helical ordering type. A detailed spin structure requires further neutron diffraction study.

The high temperature part ($T \geq 150$ K) of the $\chi(T)$ data shows a paramagnetic behavior and can be fitted with the Curie-Weiss law ($\chi(T) = \chi_0 + \dfrac{C}{T-\Theta}$) satisfactorily as shown solid red line in upper inset of Fig. 3(a). The obtained fitting parameters are $\chi_0 \cong 1.02\times 10^{-4}$ cm$^3$/(mol Cu), $C = N_A g^2 \mu_B^2 S(S+1)/3k_B \cong 0.368$ cm$^3$ K (mol Cu)$^{-1}$, and $\Theta \cong -63$ K, where $N_A$, $g$, $\mu_B$, $k_B$, and $\Theta$ represent Avogadro number, Lande-$g$ factor, Bohr magneton, Boltzmann constant, and the Curie-Weiss temperature, respectively. The core diamagnetic susceptibility ($\chi_{dia}$) of Ba$_2$CuTeO$_6$ is estimated to be $-1.12\times 10^{-4}$ cm$^3$/mol per formula unit from the corresponding ions of Ba$^{2+}$, Cu$^{2+}$, Te$^{2+}$ and O$^{2-}$.[36] The estimated Van-Vleck paramagnetic susceptibility $\chi_{vv}$, i.e., after $\chi_o$ is corrected with the core diamagnetic contribution ($\chi_{dia}$), is found to be ~$2.14\times 10^{-4}$ cm$^3$/mol, which is of similar magnitude to other cuprates.[37-39] The estimated effective magnetic moment ($\mu_{eff}$) of Cu$^{2+}$is ~1.72 $\mu_B$, which is close to the theoretical spin-only value of Cu$^{2+}$($S$ =1/2) ions ($\mu_{eff}$ = 1.73 $\mu_B$).

Below 150 K, deviation from the Curie-Weiss law occurs and develops into a broad maximum in $\chi(T)$ near $T_{max}$~ 75 K, which indicates the existence of a short-range spin correlation. Attempt has been made to fit the data to the modified Bonner-Fisher AFM chain model that include a parameter $J_{inter}$ to account for the interchain interactions,[40,41] as shown in Eq. (1) and in series expansion form of Eq. (2), which provided a satisfactory fit as shown solid blue line in main panel of Fig. 3(a). The obtained fitting parameters are $g = 2.07$, the intrachain exchange coupling $J = 48.6$ K, and the average interchain exchange coupling $J_{inter} = 22.8$ K between $Cu^{2+}$ spins:

$$\chi_{chain} = \frac{\chi_{BF}}{1 - \frac{2(J_{inter})\chi_{BF}}{N_A g^2 \mu_B^2}} \quad (1)$$

$$\chi_{BF} = \frac{N_A g^2 \mu_B^2}{k_B T} \frac{0.25 + 0.14995x + 0.30094x^2}{1.0 + 1.9862x + 0.68885x^2 + 6.0626x^3} \quad (2)$$

where $x = |J|/k_B T$.

We present our magnetization data as a function of magnetic field $H$ along the two crystal orientations $H\|ab$ and $H\perp ab$ plane in Fig. 4(a). Magnetization isotherms at 2 K with magnetic field up to 70 kOe were obtained, where no field or temperature hysteresis is observed. A clear step increase of $M(H, T = 2$ K$)$ is observed in the range near ~10 - 20 kOe for $H\perp ab$ only, as also shown in Fig. 4(b) for its derivative. The significant increase of d$M$/d$H$ (Fig. 4(b)) above the critical field strongly suggests the occurrence of a spin-flop transition, i.e., the enhanced spin susceptibility at higher field for $H\perp ab$ could be resulted from the magnetic field induced spin-flop transition so that spin direction of the AF ordered spins is flopped from its original $\perp ab$ direction to the $\|ab$ direction.[42] Moreover, d$M$/d$H$ curves [inset of Fig. 4(b)] do not show the spin-flop transition above ~15 K. These results are consistent to the proposal that a long range AF spin ordering has occurred below $T_N$~15 K and the spin anisotropy is near the $\perp ab$ direction, which agrees with the observation that $\chi(T)$ reduction below $T_N$~15 K is observed along the $\perp ab$ direction (see Fig. 3(a)).

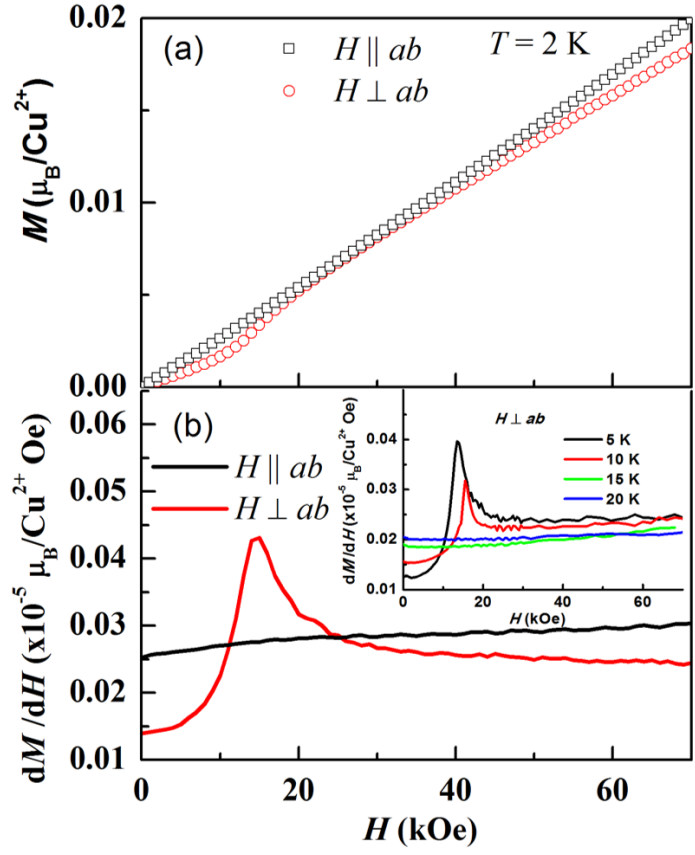

Figure 4: (Color online) (a) The field dependence of magnetization at 2 K for $H\|ab$ and $H\perp ab$, the first derivative of $M(H)$ for both directions are shown in (b) with an inset to illustrate the $dM/dH$ across the $T_N$ in $H\perp ab$.

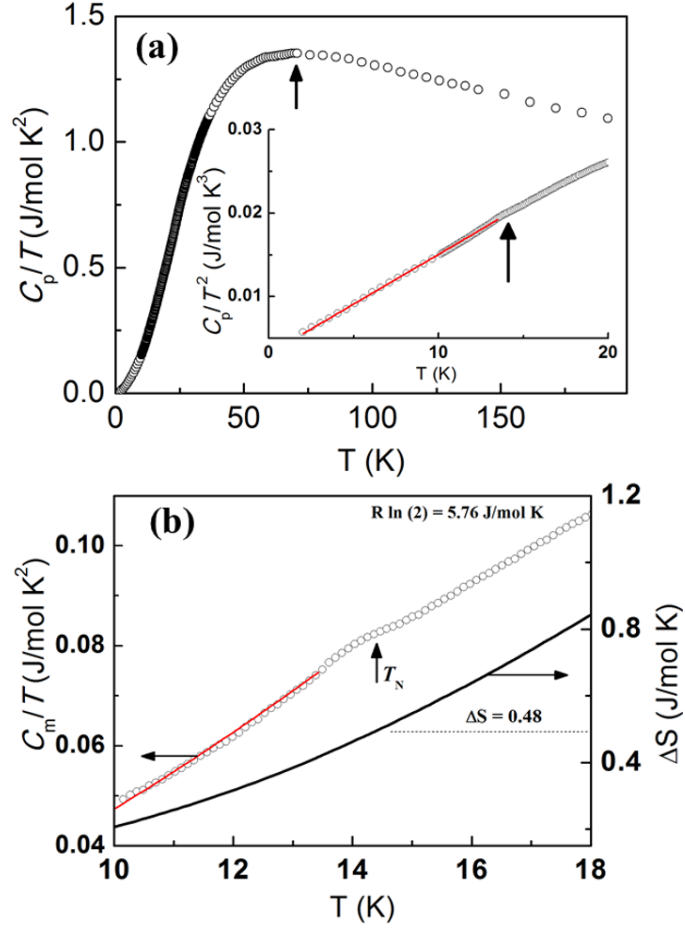

Figure 5: (Color online) (a) The temperature dependence of the total heat capacity ($C_p/T$) for $Ba_2CuTeO_6$ single crystal. Inset shows the $C_p/T^2$ vs. $T$ curve. The arrows indicating the corresponding characteristic temperatures of $T_{max}$ and $T_N$. (b) $C_m/T$ vs. $T$ and the entropy change $\Delta S$ as a function of temperature, where $C_m$ is derived from total $C_p$ with lattice contribution subtracted. The $\Delta S$ at $T_N$ corresponds to about 8% of the total Cu spin entropy of $R\ln 2 \sim 5.76$ J/(mol K) for $S = \frac{1}{2}$. The red solid line shows the heat capacities derived from the spin wave theory, as described in the text.

## C. Heat capacity

The heat capacity $C_p(T)$ measurement results for $Ba_2CuTeO_6$ single crystal at zero field are presented in Fig. 5(a). It is also found that the $C_p(T)$ data for 30 kOe coincides with the zero-field data (not shown here). As shown in the plot of $C_p/T$, no significant $\lambda$- type peak attributable to a long range magnetic ordering is found down to 2 K. On the other hand, $C_p/T$ curve exhibits a broad maximum at ~ 75 K, as shown in the Fig. 5(a), which is consistent with the $\chi(T)$ maximum near $T_{max}$~75 K. A very weak anomaly near $T_N$~15 K are identifiable in the $C_p/T^2$ plot, as shown in the inset of Fig. 5(a). Peak temperaturs identified by the broad maxima and the weak anomaly are consistent to those observed in $\chi(T)$ and $d\chi/dT$ (T) plots shown in Fig. 3.

Since the nonmagnetic $Ba_2ZnTeO_6$ compound has a different crystal structure [26], the absence of a nonmagnetic isomorphic compound for $Ba_2CuTeO_6$ does not allow a direct deduction of the lattice contribution ($C_L$) accurately. The magnetic contribution of specific heat ($C_m$) was roughly estimated using a Debye $T^3$ law approximation. The heat capacity data above $T_N$ are well fitted to $C_p/T = \gamma + \beta T^2$ with $\gamma = 24 \times 10^{-3}$ J/(mol K) and $\beta = 8.13 \times 10^{-4}$ J/(mol K). Since $Ba_2CuTeO_6$ is an insulator, it is reasonable to assign the linear term $\gamma T$ to the magnetic contribution ($C_m$), as expected for homogeneous spin - $\frac{1}{2}$ chain system [43], and the $\beta T^3$ term to the lattice contribution ($C_L$). Therefore, the magnetic contribution was calculated as $C_m = C_p - C_L$. The Debye temperature of 288.6 K can be estimated by the formula $\beta = 12\pi^4 R n/(5\Theta_D^3)$, where R, $n$, and $\Theta_D$ are the gas constant, the number of atoms per formula unit (in this case $n = 10$), and the Debye temperature, respectively. The Debye temperature is consistent with those reported for the similar Cu-based systems [44,45]. The magnitude of a small anomaly near $T_N$~15 K is found to be more pronounced in the $C_m/T$ vs. $T$ plot, as shown in Fig. 5(b). The entropy change ($\Delta S$) was calculated by integrating the $C_m/T$ as a function of temperature ($\Delta S = \int (C_m/T) dT$), as shown in Fig. 5(b). The entropy recovered at $T_N$~15 K is about 0.48 J/(mol K), which is only about 8 % of the total spin entropy of $R\ln 2 \sim 5.76$ $J/(\text{mol K})$ for the Cu $S = \frac{1}{2}$. From the estimation of $\Delta S$, it is obvious that the amount of spin entropy change at $T_N$ is too little to produce a sizable $C_m$ feature like most quasi-2D systems,[46] such as copper pyrazine (pz) perchlorate $Cu(pz)_2(ClO_4)_2$.[47]

According to the spin wave (magnon) theory, the low temperatures $C_m$ follow a $T^{d/n}$ behavior, where $d$ is the dimensionality of the magnetic lattice and $n$ is the exponent in the dispersion relation ($n = 1$ for antiferromagnets and $n = 2$ for ferromagnets).[48-50] The spin wave heat capacity of a 2D antiferromagnet is proportional to $T^2$ and a 3D antiferromagnet is proportional to $T^3$. The fittings of the $C_p/T^2$ and $C_m/T$ data shown in Fig. 5(a) and Fig. 5(b) indicate that the $C$ follows a $T^{d/n}$ dependence with fitted value of $d/n \sim 2.63$, which suggests the system could be viewed as a quasi-2D antiferromagnet with a relatively weak inter-plane coupling.

### D. Theoretical calculations

Based on the geometric parameters associated with the paths for the Te-bridged SSE routes shown in Fig. 1(c), we considered seven spin exchange-coupling parameters $J_1$ - $J_7$ summarized in Fig. 6 and Table 1. In order to find out the magnetic ground state of the system, we considered various magentic configurations possible within the supercell. Of these, we selected eight spin configurations, *i.e.*, seven AFM configurations AF1-AF7 including the AFM ground state and ferromagnetic (FM) configuration for estimating nearest-neighbor exchange-coupling parameters $J_1$ - $J_7$. The spin arrangements of these configurations can be decoded from Eqs. (3)-(10). Here, negative $J$ implies parallel spin arrangement and positive $J$ implies antiparallel spin arrangement. From the geometrical parameters, it seems that $J_2$ and $J_3$ are very similar and one may consider them to be same. The same is true for $J_6$ and $J_7$. But from our calculations, we found that these two couplings are quite different. For example, spin configuration AF1 (Eq. 4) and AF2 (Eq. 5) have opposite orientations for the spins coupled via $J_2$, $J_3$, $J_6$, and $J_7$, while the rest spin orientations are the same. However, the total energy of AF1 is 0.72 meV/f.u. lower than that of AF2 state (Tabel II). Similarly, spin configuration AF3 (Eq. 6) and AF4 (Eq. 7) differ by the spin orientation of $J_2$, $J_3$, $J_6$, and $J_7$ (the rest of the spin orientations are same), have a total energy difference of 0.19 meV/f.u. (Table II). Thus $J_2$, $J_3$, $J_6$, and $J_7$ must be distinguishable.

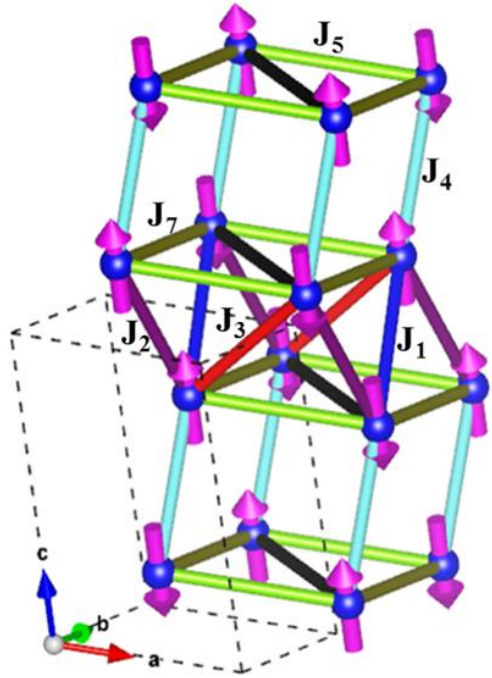

Figure 6: (Color online) Proposed spin arrangement for the configuration AF1 and exchange paths in Ba$_2$CuTeO$_6$, where different colors of cylinders represent different exchange parameters as tabulated in Table I, i.e., the major three nearest neighbor exchange constants $J_5$, $J_4$, and $J_1$ are shown in colors of green, cyan, and blue, respectively.

The relative energies of these eight spin configurations (seven AFM configurations AF1 - AF7 and one FM configuration) calculated by DFT+$U$ are summarized in Table II. We find that configuration AF1 has the lowest energy and thus we have the AF ground state consistent with the experimental finding. The magnetic moment of copper ion was found to be 0.7 $\mu_B$, slightly off to the 1 $\mu_B$ value as required by Cu$^{2+}$ oxidation state, suggesting that some of the magnetic moments lies outside the copper atomic sphere used.

Table I: Geometrical parameters of Ba$_2$CuTeO$_6$ associated with the Cu-O$\cdots$O-Cu spin exchange paths shown in Fig. 6.

| $J_{ij}$ | Cu–Cu (Å) | O$\cdots$O (Å) | ∠Cu–O$\cdots$O (deg.) | bond color |
|---|---|---|---|---|
| $J_1$ | 5.2817 | 2.6049 | 133.207;129.423 | Blue |
| $J_2$ | 5.3404 | 2.6225 | 137.021;126.321 | Purple |
| $J_3$ | 5.3439 | 2.6049 | 133.880;127.653 | Red |
| $J_4$ | 5.4616 | 2.7784 | 115.490;114.563 | Cyan |
|  |  | 2.8146 | 118.073;118.316 | Cyan |
|  |  | 2.8667 | 110.470;111.372 | Cyan |
| $J_5$ | 5.7230 | 2.6905 | 139.141;137.807 | Green |
| $J_6$ | 5.8553 | 2.8267 | 143.274;132.225 | Black |
| $J_7$ | 5.8610 | 2.7236 | 144.879;132.555 | Olive |

To extract the values of $J_1 - J_7$, we expressed the total spin exchange interaction energies of the Ba$_2$CuTeO$_6$ in terms of spin Heisenberg Hamiltonian, $H = E_0 - \sum_{\langle ij \rangle} J_{ij} \sigma_i \cdot \sigma_j$. Here $J_{ij}$ is the exchange interaction parameter between the nearest-neighbor Cu site $i$ and site $j$; and $\sigma_i$ ($\sigma_j$) is the unit vector representing the direction of the local magnetic moment at site $i$ ($j$). For AF interaction, $J < 0$ is assumed and for FM interaction, $J > 0$. The constant $E_0$ contains all spin-independent interactions.

The total energies of the supercell of all considered magnetic configurations are given by

$$E_{FM} = E_0 - 8(J_1 + J_2 + J_3 + J_4 + J_5 + J_6 + J_7) \quad (3)$$
$$E_{AF1} = E_0 + 8(J_1 - J_2 + J_3 + J_4 + J_5 - J_6 + J_7) \quad (4)$$
$$E_{AF2} = E_0 + 8(J_1 + J_2 - J_3 + J_4 + J_5 + J_6 - J_7) \quad (5)$$
$$E_{AF3} = E_0 - 8(J_1 - J_2 + J_3 - J_4 - J_5 + J_6 - J_7) \quad (6)$$
$$E_{AF4} = E_0 - 8(J_1 + J_2 - J_3 - J_4 - J_5 - J_6 + J_7) \quad (7)$$
$$E_{AF5} = E_0 + 8(J_1 + J_2 - J_3 - J_4 + J_5 + J_6 - J_7) \quad (8)$$
$$E_{AF6} = E_0 + 8(J_1 + J_2 + J_3 + J_4 - J_5 - J_6 - J_7) \quad (9)$$
$$E_{AF7} = E_0 + 8(J_1 - J_2 - J_3 - J_4 - J_5 + J_6 + J_7) \quad (10)$$

Solving the above mentioned equations, we get the values of all exchange interactions listed in Table III. It is found that the face-shared Cu-Te2-Cu dimer ($J_4$) and the two corner-shared Cu-Te1-Cu dimer ($J_1$ and $J_5$) couplings are the strongest. These AF exchange couplings are found to be $J_5/k_B$ = -45.01 K, $J_4/k_B$ = -35.12 K and $J_1/k_B$ = -30.46 K (Fig. 6 and Fig. 1(c)). $J_2$ is smallest among all the couplings and hence it is overruled by all other couplings in the ground state AF1.

Table II: Calculated total energy $\Delta E$ (relative to the total energy of FM state $E_{FM} = -56.4524$ eV/f.u.), total magnetic moment $m_s^{tot}$, atomic moment of Cu $m_s^{Cu}$.

| Config. | $\Delta E$ (meV/f.u.) | $m_s^{tot}$ ($\mu_B$/f.u.) | $m_s^{Cu}$ ($\mu_B$/atom) |
|---|---|---|---|
| FM | 0.0 | 1.0 | 0.70 |
| AF1 | -10.12 | 0.0 | 0.70 |
| AF2 | -9.40 | 0.0 | 0.70 |
| AF3 | -7.23 | 0.0 | 0.70 |
| AF4 | -7.04 | 0.0 | 0.70 |
| AF5 | -6.37 | 0.0 | 0.70 |
| AF6 | -6.10 | 0.0 | 0.70 |
| AF7 | -2.63 | 0.0 | 0.70 |

Table III: Calculated exchange interaction parameters (in K).

| $J_1/k_B$ | $J_2/k_B$ | $J_3/k_B$ | $J_4/k_B$ | $J_5/k_B$ | $J_6/k_B$ | $J_7/k_B$ |
|---|---|---|---|---|---|---|
| -30.46 | -1.10 | -4.17 | -35.12 | -45.01 | 2.63 | -2.64 |

We found that there is a large variation in the values of exchange parameters. For some set of the fitted coupling values, it varies by an order of magnitude, e.g., $J_2$, $J_3$, $J_6$ and $J_7$. At intermediate temperatures, it is expected that some of the weak couplings might be easily destroyed by the thermal fluctuations in the system. Hence the resulting magnetic structure must be dominated by the three largest couplings of $J_5$, $J_4$, and $J_1$, which strongly suggests a spin system of a spin chain system with two nearly equal interchain couplings, as indicated by the $T_{max}$ short-range spin-exchange correlation from $\chi(T)$ (see Fig. 3) and $C_p(T)$ (see Fig. 5) measurements.

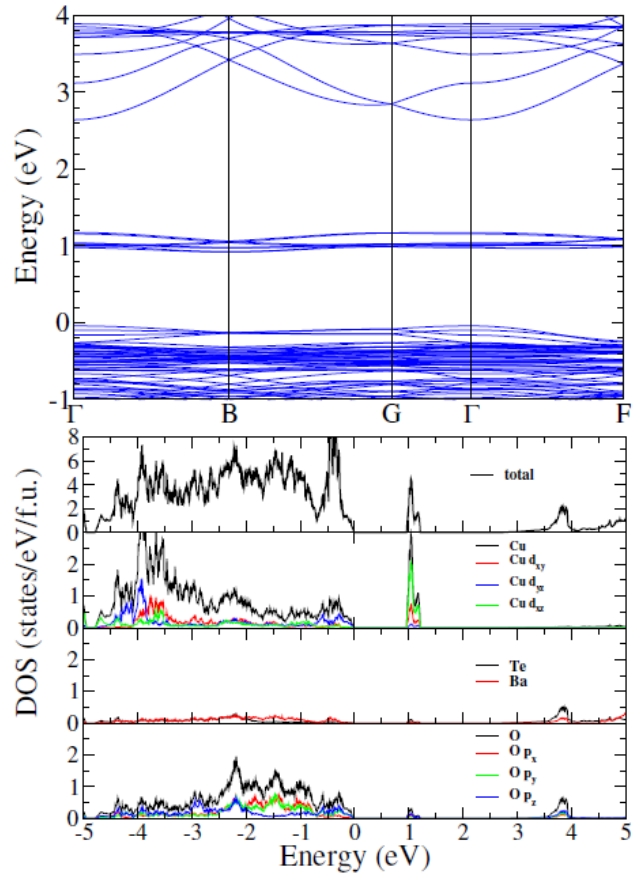

Figure 7: (Color online) Band structure (top panel) and density of states (bottom panel) of configuration AF1. Top of the valence band has been set to zero.

In Fig. 7 we plotted the band structure (top panel) and density of states (bottom panel) of configuration AF1. The conduction band has very low dispersion as it has a dominant 3$d$ character of copper atoms. From the atom-resolved density of states, it is quite clear that the valence band consists of hybridized oxygen 2$p$ ($p_z$ and $p_y$) and copper 3$d_{yz}$ states. On the other hand, the conduction band has the dominant contribution from the very narrow copper $d_{xz}$ band and a small contribution from oxygen $p_z$ and $p_x$ states. The Ba$_2$CuTeO$_6$ is an AF insulator with a band gap of 1.0 eV.

Based on the crystal structure which can be simplified as a Te-bridged CuO$_6$ octahedra, the magnetic structure is expected to be determined by the SSE coupling between the Cu spins via the Cu-O-Te1-O-Cu path for $J_5$ and $J_1$, and the Cu-O-Te2-O-Cu path for $J_4$ (Fig. 1(c)). To justify the accuracy of the theoretical values of coupling constants, we can estimate the Curie-Weiss temperature $\Theta$ in terms of the fitted coupling parameters $J_1$-$J_7$.[51] In the mean-field approximation, which is valid only in the paramagnetic limit, $\Theta$ is related to the $J_i$ as

$$\Theta = \frac{S(S+1)}{3k_B}\sum_i z_i J_i.$$

Where the summation runs over all the nearest neighbors of a given spin site, $z_i$ is the number of nearest neighbors connected by the spin exchange parameters $J_i$ and $S$ is the spin quantum number of each spin ($S = \frac{1}{2}$ for Cu atom). The calculated $\Theta$ value using the parameters obtained from GGA+$U$ calculations is -57.94 K, which agrees very well with the experimental value of -63 K from the Curie-Weiss law fitting discussed above.

### E. Te-bridged spin chain and spin dimer

The spin-chain system having two different interchain couplings has been found in many cuprate compounds experimentally, and these compounds can be viewed as a two-leg spin ladder theoretically.[52-55] The spin- 1/2 even-leg ladders are expected to have a spin-liquid ground state with short-range spin correlations [12,56,57]. On the other hand, when couplings between the ladders are not negligible, the system exhibits long-range ordering at finite temperature [58,59]. It is interesting to note that indeed the current system can also be described as a two-leg spin ladder system with intrachain coupling $J_5$ along the two legs of a spin ladder, interchain coupling $J_4$ as the rung of a two-leg ladder, and an interladder coupling $J_1$. The $J_5$ intrachain coupling has a SSE route through Te(1)-bridged $CuO_6$ octahedra via corner-sharing oxygens, $J_4$ is SSE route through Te(2)-bridged $CuO_6$ octahedra with face-shared oxygens, and $J_1$ is also a Te(1)-bridged SSE route via oxygen corner sharing as shown in Fig. 8(a). Once the 2D AF correlation is built up to certain correlation length, weak effective exchange interaction between the spin ladder planes will give rise to the AF long-range ordering of $T_N$ at low temperature, which is supported from the weak "interplane" couplings of $J_i$ (i = 2, 3, and 7) which are one order smaller than the couplings of $J_i$(i = 5, 4, and 1) responsible for the two-leg spin ladder plane, as shown in Table III and Fig. 8(b).[52,60]

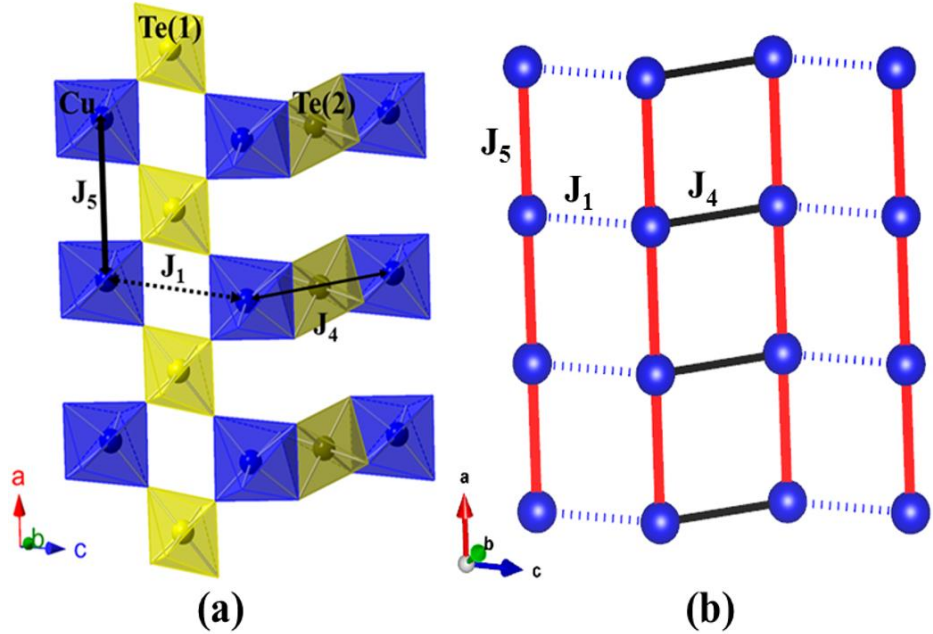

Figure 8: (Color online) (a) The layer of two-leg Cu spin ladder (see also Fig. 1(c)) formed via SSE mechanism, including the Te(1)-bridged spin chain as the leg of a ladder along $J_5$ (Cu-Te1-Cu), Te(2)-bridged spin dimer as the rung of a ladder along $J_4$ (Cu-Te2-Cu), and the Te(1)-bridged interladder coupling along $J_1$ (Cu-Te1-Cu). (b) The corresponding schematic illustration of the three parameters of a two-leg spin ladder.

The spin configuration of $Ba_2CuTeO_6$ has been implied to be a two-leg ladder based on both the structure aspect shown in Fig. 1(c) and the calculated magnetic coupling constants shown in Table III, as also compared in Fig. 8. Unlike the conventional Cu-O-Cu spin superexchange (SE) interaction discussed in the high-$T_c$ cuprate systems, all $Cu^{2+}$ spins are coupled via a super superexchange (SSE) route through either Cu-O-Te1-O-Cu or Cu-O-Te2-O-Cu paths (Fig. 1(b)). It is also interesting to note that structurally the Cu-O-Te1-O-Cu path corresponds to the oxygen corner-sharing among $CuO_6$ and $Te(1)O_6$ octahedra, on the other hand, the Cu-O-Te2-O-Cu path corresponds to face-sharing octahedra between the two, as shown in Fig. 1(c). The strength of a spin-exchange interaction between two adjacent $Cu^{2+}$ ions through SSE paths has been shown depending mostly on the orbital overlap of the two O-2$p$ orbitals and the two Cu-$d$ orbitals.[61,62] These orbitals overlap seems to increase with larger ∠Cu-O···O bond angles and shorter O···O distances, especially must be shorter than the van der Waals radii sum ~3.04 Å.[63, 64] The reason that $J_5$ is the strongest (see Table III) can be understood from the fact that O$_5$···O$_6$ bond length of 2.6905 Å is significantly shorter than the van der Walls distance, and both ∠Cu-O···O bond angles of 137.807 and 139.141° are larger comparing to those of $J_4$ and $J_1$ with the SSE path. On the other hand, even $J_1$ has a shorter O···O distance and larger ∠Cu-O···O bond angles comparing with those of $J_4$, the SSE path of $J_4$ corresponds to the face-shared CuO$_6$-TeO$_6$ octahedra having three O···O routes of bond distances: 2.7785 Å, 2.8146 Å, and 2.8667 Å, the overlap integrals of $J_4$ are stronger than that of the path $J_1$.[65] While modified Kanamori-Goodenough rules for the SSE mechanism have been proposed by Whangbo *et al.* with some verified examples and supported partly by the current *ab initio* calculations,[61,62] it is clear that the number of O···O paths, i.e., oxygen corner sharing or face sharing, must be taken into account in addition to the rules generated earlier, i.e., considering only on the O···O distance and ∠Cu-O···O bond angles for the spin SSE mechanism.

Based on the current calculations, the strongest spin exchange interaction $J_5$ supports the spin chain formation as the leg of a two-leg spin ladder, and the second strongest $J_4$ ($J_4$/$J_5$= 0.78) could be viewed to support the rung of a two-leg spin ladder, and $J_1$ ($J_1$/$J_5$= 0.67) introduces a frustrating interladder AF coupling which is slightly weaker than that of $J_4$, as shown in Fig. 8(b). This two-leg spin ladder with significant interladder couplings forms a plane and is responsible for the observed broad maximum near $T_{max}$ due to the spin-exchange coupling in short range due to the low dimensionality. At temperatures below $T_N$ of less thermal fluctuation, the much weaker AF interlayer couplings of $J_i$ (i = 7, 3, and 2) could induce the observed AF long-range orderings, as observed in both $\chi(T)$ (see Fig. 3) and $C_p(T)$ (see Fig. 5).

## IV. Summary and conclusion

The crystal growth, magnetic and thermodynamic properties of $Ba_2CuTeO_6$ with triclinic symmetry are reported. The observed experimental and theoretical results provide the picture of a two-leg spin ladder system that evolves from a short-range intrachain spin interactions and to the long range ordered 3D AF ordering in steps. In particular, the Cu spins are of SSE coupling mechanism via Cu-O-Te-O-Cu route. The magnetic susceptibility $\chi(T)$ data and its derivative $d\chi/dT$ show signatures of spin-exchange coupling of short-range AF nature at $T_{max}$~ 75 K in $\chi(T)$, before the system is ultimately driven into a 3D LRO below $T_N$=15 K. The isothermal magnetization for $H \perp ab$ plane reveal a spin flop transition with $H$~15 kOe to confirm the existence of a 3D AF LRO below $T_N$ with spin anisotrophy along the $c$ direction. The heat capacity $C_p$ of the $Ba_2CuTeO_6$ single crystal is also found consistent to the proposed spin structure of a two-leg spin ladder with significant interladder coupling in 2D. A neutron diffraction study to the title compound is underway to solve the AF spin structure below $T_N$. This study is valuable to the understanding of both spin ladder and spin coupling in SSE route.

## Acknowledgment
F.C.C. acknowledges support provided by the Ministry of Science and Technology, Taiwan (MOST-104-2119-M-002-028 -MY2). G.Y.G. thanks support from the Ministry of Science and Technology, National Taiwan University and Academia Sinica in Taiwan.